\documentclass[twocolumn, trackchanges]{aastex701}
\usepackage{multirow}
\usepackage{gensymb}
\usepackage{amsmath}

\begin{document}


\title{Exploring the origin of slow-rotating pulsars: Emission from STARS simulated white dwarfs}

\author[orcid=0009-0000-6980-6334]{Zenia Zuraiq}
\affiliation{Department of Physics, Indian Institute of Science,
Bangalore 560012, India}
\email[show]{zeniazuraiq@iisc.ac.in}

\author[orcid=0000-0002-3020-9513]{Banibrata Mukhopadhyay}
\affiliation{Department of Physics, Indian Institute of Science,
Bangalore 560012, India}
\email[show]{bm@iisc.ac.in}

\begin{abstract}

An increasing number of slow-rotating pulsars and long-period transients have been observed in recent years. The periods of these sources can be significantly longer than typical pulsar sources, ranging from $\sim 100 - 1000$ s, or even longer. The observed periods and period derivatives predict high magnetic fields, higher than the Schwinger field, when we consider conventional pulsar models of dipole-driven emission from neutron stars. Interestingly, the periods of these sources are consistent with the current observations of white dwarfs (WDs). We explore the possibility of magnetic WDs, simulated using the stellar evolution code \texttt{STARS}, as the origin of some of these unusual pulsar observations. We simulate WDs with enhanced magnetic fields due to accretion after their formation. We also briefly touch on the more exotic possibility of super-Chandrasekhar WDs as the origin for these sources. We find that expanding the parameter space to include the possibility of a WD origin 
provides an alternate and viable explanation for these unusual observations.

\end{abstract}



\section{Introduction} 
\label{sec:intro}

Long period transients (LPTs) are an intriguing and exciting class of objects that are being increasingly observed currently \citep{LPTreview}. These are a new class of radio sources which have rotational periods from minutes to hours and, hence, are much longer than conventional pulsar sources. There are at least 12 LPT sources that have been detected till today. In general, pulsar radio emissions are mostly understood to be due to the conversion of rotational energy into dipolar emission from a rotating, magnetized neutron star \citep[NS; ][]{LK2004}. However, the high periods of LPTs are not easily explained based on NS emission ideas. In fact, they often violate the conventional death valley (e.g. \citealt{CR1993}) idea of pulsar emission. It is apparent that such sources are then a way to probe the pulsar emission mechanism and our pre-conceptions of the same in more detail.

It appears that LPTs are a diverse class and may not all have the same origin. As mentioned above, a conventional rotating dipole NS model turns out to be insufficient for the present purpose. In fact, when estimating the magnetic field required to power their emission in the rotating dipole scenario, the fields often end up being higher than the critical (Schwinger) field, $B_c = 4.41 \times 10^{13} \ G$, a regime where pair production, and thus pulsar emission, is believed to be suppressed \citep{BH1998}. 

A class of objects that seems to bridge the gap between the conventional NS pulsars and the LPTs is the slow-rotating pulsars\footnote{Also called long period pulsars/ultra-long period pulsars/ultra-long period magnetars in the literature. } (SRPs). This includes one of the slowest rotating pulsars, J0901-4046~\citep[period, $P=75.88 ~s$; ][]{caleb}, which itself requires $B > B_c$ under conventional NS dipole models. Interestingly, there are at least two sources observed currently, that are confirmed white dwarf (WD) pulsars. These are  AR Scorpii~\citep[$P=117.12 ~s$;][]{arsco} and J1912-4410~\citep[$P=319.3~s$;][]{j1912}. Further, the position of J0901-4046 in the $P - \dot{P}$ diagram, where $\dot{P}$ is the period derivative, is very close to AR Scorpii. 

The uncertain nature of LPTs/SRPs, along with their similar $P$ and $\dot{P}$ (at least in the case of J0901-4046) to confirmed WD pulsar sources, makes the idea of WDs as the origin of these sources a serious possibility worth investigating. In the case of AR Scorpii, the pulsation is thought to be driven by binary interactions with an M-dwarf companion, although the exact emission model is still unclear \citep{arsco_Buckley}. Alternatively, the emission may also be driven by the synchronization of the WD spin with
the orbital motion in a polar-like system, leading to the “synchronar” model \citep{katz_synchronar}. Similar WD-binary interaction or connections have been confirmed for two LPTs, i.e., ILT J1101+5521~\citep[$P=123.53 ~\mathrm{min}$;][]{ilt,zhang_ilt} and GLEAM-X J0704-36~\citep[$P=174.94 ~\mathrm{min}$;][]{gleamxj0704_1,gleamxj0704_2}, and plausible WD binary origins are suggested for ASKAP J1935+2148 \citep[$P=53.76 ~\mathrm{min}$;][]{askap} and GPM J1839-10 \citep[$P=1318.19~s$;][]{gpm}, with less stringent constraints. Additionally, ILT J163430+445010 ($P=841.25~s$), the first LPT to be linked with a possible double WD system, was detected recently \citep{bloot}. An AR Scorpii like WD binary interaction model appears to be a viable explanation for some of the LPTs observed today.

However, as proposed previously \citep{katz_gleamx,GilZhang05}, \textit{isolated} WDs may also produce magnetospheric emission, similar to typical NS pulsars. This is particularly useful in explaining the shorter period of LPTs whose periods range in $\sim 100 - 1000 \ s$, similar to the observed period of rotating WDs. These can fit the cases of LPTs without evidence for binary companions, and even SRPs like J0901-4046. However, if a small number of LPTs/SRPs are isolated WDs, the question then arises: why are isolated WD pulsars so rare, when compared to the regular WD population?  As we explore further in this work, it appears that only the highly magnetized WDs, with surface fields $\gtrsim 10^8 \ G$, can behave as sources adequate to explain the pulsar nature of LPTs/SRPs, which are a rare subset of the isolated WD population \citep{ferrario2020}.

We aim to tackle the possible WD origin of LPTs/SRPs by using models of simulated magnetized WDs derived from time-dependent stellar evolution. For this, we use the Cambridge stellar evolution code \texttt{STARS} \citep{eggleton}. From stable stellar evolutionary tracks, we get realistic WD masses, radii and magnetic fields, which can then serve as the origin of pulsar emission. 

We simulate the evolution of WDs from magnetized main sequence stars (MSs). We consider the stars to have density-dependent magnetic fields that vary from their center to surface. The details are explained in Sec. \ref{sec:formalism}. We further simulate the accretion of material onto this WD from a companion after its formation, leading to its contraction and enhanced magnetic fields \citep{ZuraiqApJL}. This enhanced magnetic field can also lead to enhanced pressure support, leading to the exotic case of massive WDs that exceed the Chandrasekhar mass-limit \citep{ZuraiqApJL,deb_wd,DB2013,SSM2015}. We explore the possibility of super-Chandrasekhar WDs as the origin of LPT/SRP pulsar emission in Sec. \ref{sec:SCWD}.

Thus, accretion is one of the possible routes towards highly magnetized and, perhaps, highly massive WDs. We explore whether the end points of such an evolution leading to highly magnetized WDs can serve as the origin of pulsar emission. Importantly, the WDs need not necessarily be in a binary system today, and could also be an isolated WD, but with a history of accretion from a companion. We have reported our detailed exploration of the enhancement of magnetic field and/or mass due to accretion elsewhere \citep{ZuraiqApJL}.

\subsection{Sources under consideration}
The sources we target, along with their properties, are given in Table~\ref{wdp_table}. We include the SRP J0901-4046, which, as mentioned previously, has $P$ and $\dot{P}$ very similar to the confirmed WD pulsar AR Scorpii. We also include GLEAM-X~J162759.5-523504.3 \citep[GLEAM-X J1627 for short; ][]{gleamx} and the longest active LPT, GPM J1839-10 \citep{gpm}, which have periods $\sim 1000~s$. Recent work has examined the possibility of these sources, in particular GPM J1839-10, being associated with a WD in a binary, with the binary companion being  an evolved M-dwarf, leading to an AR Scorpii-like scenario \citep{gpm_nature}. Isolated WD models have been explored for GLEAM-X J1627 \citep{dwd_gleamx,katz_gleamx}.

\begin{table}[h]
    \centering
    \begin{tabular}{lll}
    \hline
       Source  & $P/{\rm s}$ & $\dot{P}/\rm s\,s^{-1}$ \\     \hline
        (a) J0901-4046 & 75.88  &  $2.25 \times 10^{-13}$\\
        (b) GLEAM-X J1627 & 1091.17  &  $6 \times 10^{-10}$\\
        (c) GPM J1839-10 & 1318.19  &  $3.6 \times 10^{-13}$\\
        \hline
    \end{tabular}
    \caption{Observed properties of some SRPs/LPTs.}
    \label{wdp_table}
\end{table}

In the current work, we examine if the properties of realistic simulated magnetized WDs, arising from stellar evolution, can explain these sources. Particularly, we focus on if the mass ($M$), radius ($R$) and magnetic field ($B$) can explain the observed $P$-$\dot{P}$ of the SRPs/LPTs listed in Table~\ref{wdp_table}. The obliquity angle $\alpha$, i.e. the angle of misalignment between magnetic and rotation axes, ends up being a free parameter in this study. On a parallel track, we explore the evolution of the star by a theoretical/semi-analytical WD evolution based on the conservation of momentum and magnetic flux \citep{BM2017} in our companion paper (Mukhopadhyay \textit{et al.}, Paper I), particularly for the case of J0901-4046, and obtain similar results and constraints as we present here.

The rest of the paper is organized as follows. In Sec. \ref{sec:formalism}, we describe our formalism for the stellar evolution simulations using the \texttt{STARS} code. In Sec. \ref{sec:alpha}, we report our results of combining the simulated WDs with observed slow-rotating sources. We discuss the more exotic scenario of massive, super-Chandrasekhar WDs in Sec. \ref{sec:SCWD}. We end with our conclusions in Sec. \ref{sec:conclusions}.

\section{\texttt{STARS} Simulation set-up and formalism} 
\label{sec:formalism}

We have simulated the formation and evolution of magnetized WDs starting from magnetized MSs using the Cambridge stellar evolution code, \texttt{STARS} \citep{eggleton, zz_stars}. We then examine \texttt{STARS} results under the standard pulsar emission model to probe the nature of certain slow-rotating sources. 

\texttt{STARS} is a one-dimensional (1D) stellar evolution code that solves for the following stellar structure equations,
\begin{equation}
    \frac{d\log_e P}{dm} = -\frac{Gm}{4\pi r^4P},
\end{equation}

\begin{equation}
    \frac{d\log T}{dm} = \frac{d\log P}{dm}\nabla,
\end{equation}

\begin{equation}
    \frac{d \log_e r}{dm} = \frac{1}{4\pi r^3m},
\end{equation}
and
\begin{equation}
    \frac{dL}{dm} = \epsilon - \epsilon_\nu - \frac{Du}{Dt} + \frac{P}{\rho^2}\frac{D\rho}{Dt},
\end{equation}
where $m$ is the mass at radius $r$ in a star at time $t$, and $T$, $L$, 
$P$, $\epsilon$, $\epsilon_\nu$, $u$, $\rho$ and $G$ are the corresponding
temperature, luminosity, pressure, energy generation rate by nuclear reactions,
neutrino energy loss rate, internal energy per unit mass, density and gravitational constant respectively.  

The logarithmic
temperature gradient with pressure, $\nabla$, depends on whether the star
is locally convective or radiative. This is defined in terms of the quantities $$\nabla_a = \left(\frac{\partial \log{T}}{\partial \log{P}}\right)_{s}$$ and $$\nabla_r = \frac{3\chi PL}{16\pi a cGmT^4},$$
where $\chi$ is the opacity, $a$ is the radiation density constant, and c is the speed of light. 
In radiative regions 
($\nabla_r \leq \nabla_a$), 
\begin{equation}
\nabla = \nabla_r.
\end{equation}
In convective regions ($\nabla_r \geq \nabla_a$), 
\begin{equation}
\nabla = \nabla_a + F(\nabla_r - \nabla_a),
\end{equation}
where $F$ is fixed based on the theory of convection considered. In the \texttt{STARS} code, the  B{\"o}hm-Vitense mixing length theory \citep{BV_mlt} is followed. See the original \texttt{STARS} papers, \citealt{eggleton,stars1}, for more details.

We solve for a network
of 21 nuclear reactions that are important for stellar structure. See previous work \citep{eggleton,stars1,stars2,stars3} for the details of the original \texttt{STARS} code, as well as the following years of updates.

\subsection{Introducing magnetic fields}

\subsubsection{Field profile and geometry}

We introduce magnetic fields in \texttt{STARS} following a density-dependent profile \citep{bandyopadhyay}, used in previous WD studies \citep{ZuraiqApJL,zz_stars,deb_wd}, given by,
\begin{equation}
\label{prof}
B(\rho) = B_{\rm s} + B_0\left[1 - \exp\left\{-\eta{\left(\frac{\rho}{\rho_0}\right)^\gamma}\right\}\right],
\end{equation}
where $B_{\rm s}$ is the surface field, $B_0$, $\eta$, $\rho_0$ and $\gamma$ are
model parameters that determine the strength and profile of the
magnetic field from the centre of the star to its surface. We fix  $\rho_0 = 10^9$ g/cm$^3$ throughout the current work. During stellar evolution, as the star contracts or expands, the magnetic field also correspondingly increases or decreases. This follows the behaviour of approximate flux freezing. 

As a result of the magnetic field within the star, the pressure is modified as $P = P_m  + P_B$, where the total pressure $P$ has contributions from both matter pressure ($P_m$) and magnetic pressure ($P_B$). Thus, including  eq. \eqref{prof} within \texttt{STARS} results in an evolving magnetic field, following the stellar evolution of the star, with magnetic fields incorporated in the stellar structure through the magnetic pressure contribution.

It has long been known that purely poloidal or purely toroidal stellar magnetic fields result in instabilities \citep{T1973,MK1973}. We thus expect that realistic WDs (and stars in general) have a mixed field geometry. Further, from previous stellar evolution studies, it is expected that stars have toroidally dominated magnetic fields at the end of main sequence \citep{quentin}. This is also seen from stability studies \citep{BR2009}. 

We approximate different field geometries by considering radially oriented (RO) and transversely oriented (TO) fields. These are the 1D equivalents of poloidal and toroidal fields respectively, leading to $P = P_m \pm B^2/8\pi$. For RO fields, $P_B = -B^2/8\pi$, while TO fields have $P_B = B^2/8\pi$. This follows from the magnetization tensor, i.e., $M_{ij}~=~(B^2/8\pi)\delta_{ij}~-~B_iB_j/4\pi$, where $B_i$ are the components of the magnetic field in Cartesian coordinates. In the case of RO fields, $\textbf{B} = (B_r, 0, 0)$ and for TO fields, $\textbf{B} = (0, B_\theta, B_\phi)$. In the stellar structure equations, the total magnetic term turns out to be $\nabla_jM_{ij} = \nabla_i(B_j^2/2) - B_j \nabla_jB_i$. Consider the examples of  RO field $\textbf{B} = (B_r, 0, 0)$ and the TO field $\textbf{B} = (0, 0, B_\phi)$. For the former, $M_{ij}$ = diag($-B^2/8\pi$, $B^2/8\pi$, $B^2/8\pi$) in matrix form.
Thus, it is clear that $M_{rr} = -B^2/8\pi$ for the RO case. Hence the surface field having poloidal geometry exhibits $P=P_m-B^2/8\pi$. Similarly one can get the $M_{ij}$ form for the above-mentioned purely azimuthal TO field as diag($B^2/8\pi$, $B^2/8\pi$, $-B^2/8\pi$), leading to $M_{rr} = B^2/8\pi$. In case of a TO field in an arbitrary direction, i.e., $\textbf{B} = (0, B_\theta,B_\phi)$, the $M_{\theta\theta}$ and $M_{\phi\phi}$ terms change without affecting the $M_{rr}$. This confirms that TO fields exhibit $P=P_m+B^2/8\pi$ in the $r$ direction. See \citealt{deb21}, for more details.

In order to implement approximately stable mixed field geometries in our simulation, we consider the inner $85\%$ of the star to have a toroidal/TO field: $P = P_m + B^2/8\pi$. The outer portion of star has a poloidal (RO) field, with $P = P_m - B^2/8\pi$, corresponding to the observed (mostly) dipolar fields from the surfaces of stars.

Previous work \citep{zz_Massgap} showed that it is not just the magnitude of the magnetic field, but also its profile, i.e., how it varies from center to surface, that plays an important role in stellar structure. In the present work, we consider $\gamma = 0.8$ with $\eta = 1.1, 1.3$.

\subsubsection{Field evolution and flux freezing}

The central magnetic field of the WD is controlled by the parameter $B_0$, which we fix to $10^{14} \ G$ throughout this work. Note that the central magnetic field is also controlled by the central density - thus, it is much lower in the MS stage ($\sim 1 - 10$ g/cm$^3$), whereas it strengthens as we approach the degenerate WD stage, due to the resultant contraction. This follows approximate flux freezing. In the WD stage, we obtain central magnetic fields $\simeq 10^{11} - 10^{12} \ G$. As the WD accretes matter, as discussed in the following subsection, its central field further strengthens and can reach $\lesssim B_0$. The maximum central field in the WDs considered in the present work is $6 \times 10^{13} \ G$. The corresponding central densities may be as high as $10^9$ g/cm$^3$, while before accretion starts, it could be as low as $\sim 10^6$ g/cm$^3$. In terms of stellar structure and pressure balance, it is the central field which plays a cruicial role. However, in the present context, when comparing with pulsar observations, it is the surface field which is key, as this is what drives emission. 

We fix the surface fields of the stars through explicit flux freezing, i.e., $B_sR^2$ being constant. This is justified by the near isothermal matter inside the WD. We consider different constants, i.e., $B_sR^2 = 5 \times 10^4 \ GR_\odot^2$, $B_sR^2 = 10^5 \ GR_\odot^2$, and $B_sR^2 = 5 \times 10^5 \ GR_\odot^2$, corresponding to different levels of magnetization in the system. These correspond to roughly the magnetic fluxes observed from the highly magnetized Ap/Bp stars in the main sequence \citep{W2014}. The resultant final surface fields of the evolved and simulated magnetized WD is consistent with the highest fields seen in observations, i.e. $B_s \sim 10^8 - 10^{10} \ G$. \citep{ferrario_wd,ferrario2020}. 

\subsection{Simulation results}

We start with a zero age main sequence (ZAMS) star that undergoes nuclear burning and normal stellar evolution till the point of exhaustion of hydrogen fuel in its core. The star then moves to the asymptotic giant branch (AGB) phase, during which time the degenerate core grows within the star. We simulate the loss of outer layers/envelope of the star at this stage through a mass loss/stellar wind control in the code, with the exact mass loss rate set within the typical values found in literature \citep{agb_mloss}. A partially degenerate star is left behind after this, which further evolves and contracts to form a mostly degenerate WD, with a thin non-degenerate envelope. The exact composition of the final WD is determined by the initial model and subsequent mass loss in the star. 

\begin{figure}[!h]
	\centering
    \includegraphics[scale=0.25]{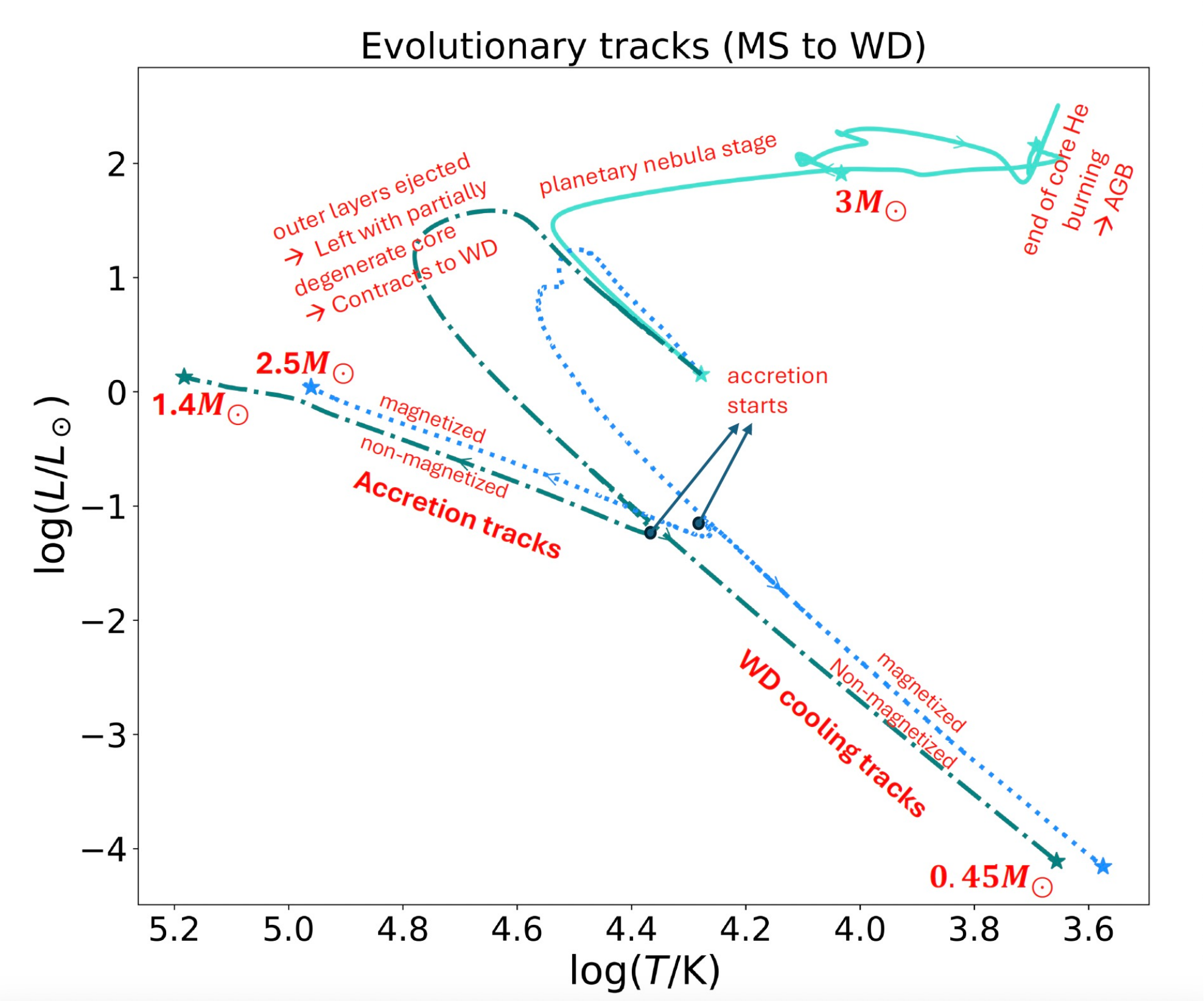}
	\caption{{Evolution of MS star to WD in a H-R diagram. The initial part of the evolution (solid line) is the evolution of $3M_\odot$ MS star leading to the AGB stage, which is not affected by the magnetic field. The subsequent evolution depends on whether the star is magnetized (dotted) or non-magnetized (dot-dashed). Accretion tracks from the magnetized and non-magnetized cases, as well as the resultant mass limit(s), are also shown.}}
    \label{HR}
\end{figure}

Our aim, in the present work, is to simulate the stellar evolution of magnetized MSs as they evolve through the various stages of stellar evolution and form magnetized WDs. We visualize the evolution through the underlying Hertzsprung-Russell (H-R) diagram, which now has information of the magnetic field, as shown in Fig.~\ref{HR}. We show the evolution of a $3M_\odot$ MS to a $0.45M_\odot$ HeCO WD \citep{heco}, passing through the AGB. The cooling tracks are slightly different for magnetized and non-magnetized cases, though it does not affect the mass.

\begin{figure}[!h]
	\centering
    \includegraphics[scale=0.22]{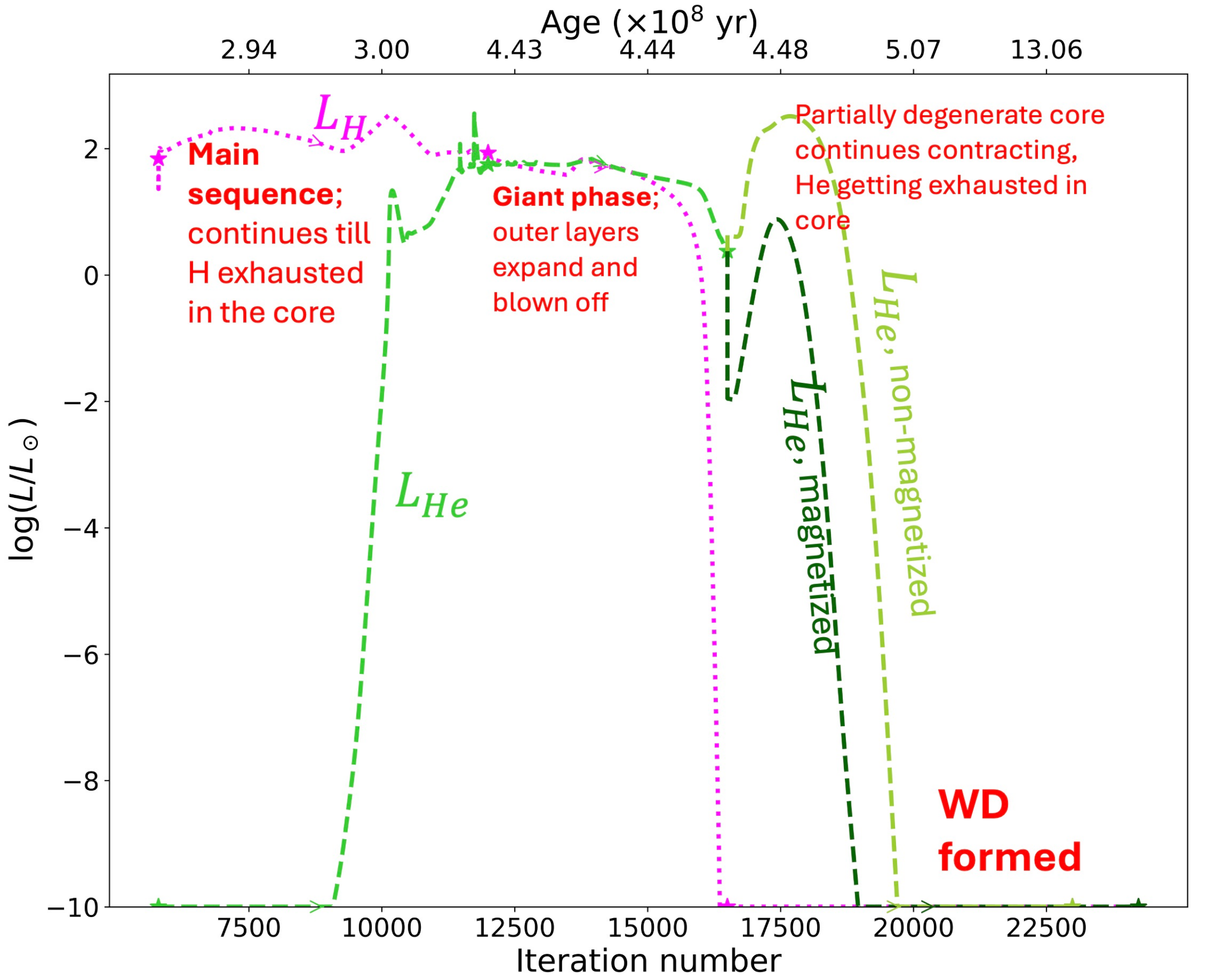}
	\caption{{Evolution of luminosities of hydrogen ($L_H$) and helium ($L_{He}$) during the different stages of evolution from MS to WD. These track the fusion of hydrogen and helium respectively during stellar evolution. Note the slight difference in magnetized and non-magnetized tracks for $L_{He}$.}}
    \label{lum}
\end{figure}

We can also track the stellar evolution from MS to WD in the \texttt{STARS} code based on the luminosities of hydrogen and helium during the various stages of stellar evolution, as shown in Fig. \ref{lum}. We see that the helium luminosity increases steadily in the initial part of the evolution, as more and more helium is accumulated in the core from the hydrogen fusion. As the star leaves the main sequence, all the hydrogen in the star is confined mainly to the outer layers of the star, as nuclear fusion exhausts hydrogen in the core. This is where we introduce a wind through the \texttt{STARS} mass loss control to mimic the mass loss during the giant/AGB phase. The star's outer layers expand and this is reflected in the hydrogen and helium luminosities going down. Eventually, the outer layers are blown off the star in a planetary nebula (modeled by the stellar wind) and we see the hydrogen luminosity goes down to zero (note that in this particular code, $10^{-10}$ is equivalent to zero). Simultaneously, the core contracts and helium ignition happens in the core, leading to the formation of a carbon-oxygen (CO) core. Thus, helium is also steadily getting exhausted in the core, although for the current scenario, shown in Fig. \ref{lum}, there is still some helium in the outer core, leading to a HeCO WD. Once the mass loss phase is done, any remaining helium in the core of the star undergoes nuclear fusion, while the core further contracts. This eventually leads to the formation of a degenerate CO core, and thus the helium luminosity goes down.

\begin{figure}[!htb]
     \centering
         {\centering
         \includegraphics[width=0.4\textwidth]{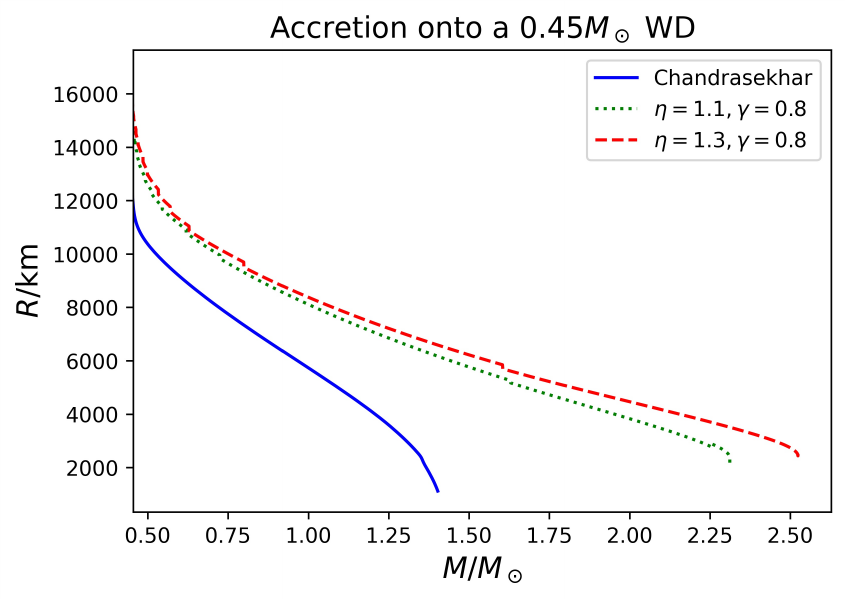}

     \vskip\baselineskip
            \includegraphics[width=0.4\textwidth]{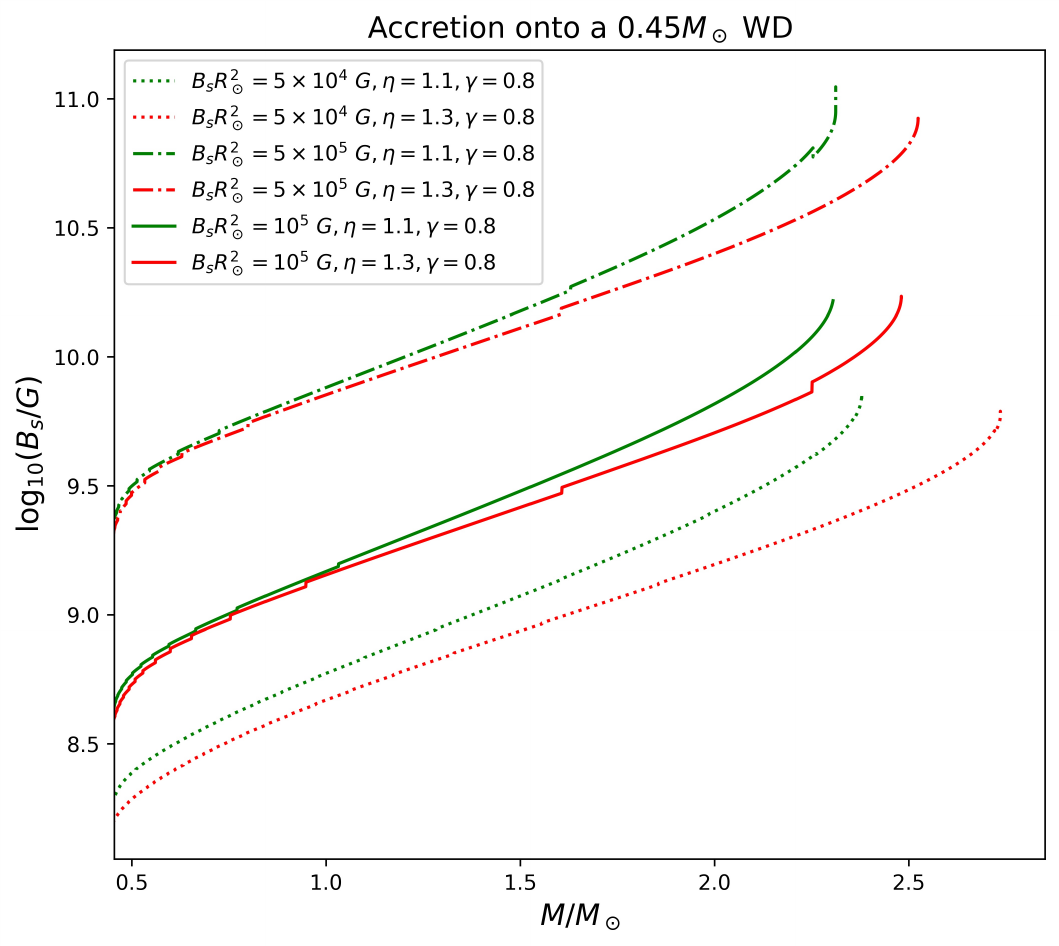}}

\caption{{\textit{Upper: }The $R-M$ relation as a result of accretion on to a $0.45M_\odot$, shown for non-magnetic (Chandrasekhar) case and magnetic cases of two different profiles. \textit{Lower: } $B_s-M$  relation showing the strengthening of $B_s$ due to contraction of the WD by accretion. The saturated $B_s$ depends on the initial magnetic flux, $B_sR_\odot^2$.}}
    \label{RMBM}
\end{figure}

In the initial part of the stellar evolution, i.e., MS~$\rightarrow$~WD formation, we find that the magnetic fields are largely \textit{dormant}, i.e., they do not affect the stellar structure. This is seen in Fig.~\ref{HR}, as the initial evolution, shown with solid line, is the same in the presence and absence of the magnetic field. Dormantly magnetized WDs, however, can undergo further evolution through accretion, resulting in the strengthening of their magnetic fields. This corresponds to the differing magnetic (dotted) and non-magnetic (dot-dashed) evolution shown in the H-R diagram of Fig.~\ref{HR}. The $B_s - M$ and the $R-M$ relations as a result of accretion of matter onto a 0.45$M_\odot$ WD are shown in Fig.~\ref{RMBM}. Note that $B_s$ does not affect the limiting mass, this is controlled by the central field of the WD. Similar strengthening of $B_s$ and resultant higher mass limit(s) can also be found in the theoretical model of accretion-based evolution explored in Paper I.

Due to the additional magnetic pressure, magnetized WDs (particularly with TO fields) can support higher masses than the Chandrasekhar limit. This is seen in the $R-M$ relation of Fig.~\ref{RMBM}. The possibility of super-Chandrasekhar WDs has been studied in theoretical and semi-analytic studies previously~\citep{DB2013,SSM2015,deb_wd,zz_stars}. We investigate the possibility of super-Chandrasekhar WDs in the current context of SRPs/LPTs in Sec.~\ref{sec:SCWD}. Further details regarding the \texttt{STARS} simulation of stellar evolution, and their link to current observations, are discussed in detail elsewhere~\citep{ZuraiqApJL}.

Thus, in the present work, we work with a simulated collection of different magnetized WDs, formed from stellar evolution of a 3$M_\odot$ MS to a $0.45M_\odot$ dormantly magnetized WD, and subsequent accretion on to it. The accretion history of the WD formation is not particularly pertinent in the present study, and these could be magnetized WDs formed as the end-stages of other stellar evolution mechanisms as well. The primary role of accretion is in strengthening the magnetic field and subsequently, $B_s$ of the WD, allowing us to simulate a collection of WDs, of different $M$ and $R$, and with $B_s$ corresponding to their highest observed fields currently.

Based on the estimates from the observations of confirmed WD pulsars AR Scorpii and J1912-4410 \citep{BuckleyNatA,arsco_surf1,j1912_mag}, the observational $B_s \simeq 10^8 \ G$. Further, in previous work, the non-resonant inverse Compton scattering (ICS) was shown to be expected as the most efficient mechanism in WD magnetospheres \citep{GilZhang05, Zhang1997}. These require effective fields $\sim 10^9 \ G$. Hence, we study those simulated pulsars with $B_s \approx 10^8 - 10^{10} \ G$. These are on the higher side of observed WD fields \citep{ferrario_wd,ferrario2020}, as expected from fossil field evolution from the most magnetized main sequence Ap/Bp stars \citep{W2014}. 

\subsection{Justifying our approach}
We undertake a full stellar evolution simulation, incorporating realistic nuclear reaction networks and magnetic field, as a MS evolves to a magnetized WD, passing through AGB. We then investigate these simulated WDs to check their feasibility as SRPs/LPTs. In the process, we make several simplifying assumptions. We discuss these below along with their justifications. 

\begin{enumerate}
    \item \textit{Absence of coupled Maxwell equations}: In the present code, we do not solve the Maxwell equations explicitly. However, our chosen magnetic field profile, given in eq. \eqref{prof} gives us only the magnitude of the field. This magnitude can be split into components, that are consistent with the Maxwell equations under our current assumptions, i.e. spherical symmetry, for both RO and TO fields. Thus, in our present 1D context, our magnetic star solutions are consistent with Maxwell's equations. A detailed proof of the same is given by  \cite{deb21}.
    \item {\textit{Magnetic fields in a 1D code}}: As mentioned previously, we simulate stars with dominant toroidal fields, as expected from the end of stellar evolution from simulations \citep{quentin} and stability considerations \citep{BR2009}. It has been shown previously from realistic two-dimensional simulations, using the general relativistic magnetohydrodynamical code XNS, that toroidal fields result in minimal deviation from spherical symmetry \citep[$\lesssim 10\%$;][]{SSM2015}. For a first approach/first-order study of the effect of magnetic field on stellar evolution, this margin of error can be neglected. 

    \item{\textit{Stability of magnetic field/profile}: The magnetic field profile we use is physically motivated, and has been used in several past studies of compact objects, including WDs \citep[e.g.][]{deb_wd}. Further, our magnetic field variation from center to surface is consistent with the exploration of axisymmetric two-dimensional Einstein-Maxwell solutions based on codes like \texttt{XNS}   \citep{SSM2015,franzon_wd}.  Importantly, as we presently perform time-dependent simulations, we find that we cannot include arbitrarily high field magnitudes and/or arbitrarily steep magnetic field profiles, as this leads to the destabilization of the star and halting of further evolution. The cases reported in the present work are those corresponding to complete evolution, where this destabilization does not occur. Hence the magnetic field magnitudes and profiles used are those that ensure stable evolution. A few representative magnetic field profiles, i.e. the variation of magnetic field within a given WD are given in Fig. \ref{Bprofs}.}

    \begin{figure}[!h]
	\centering
    \includegraphics[scale=0.6]{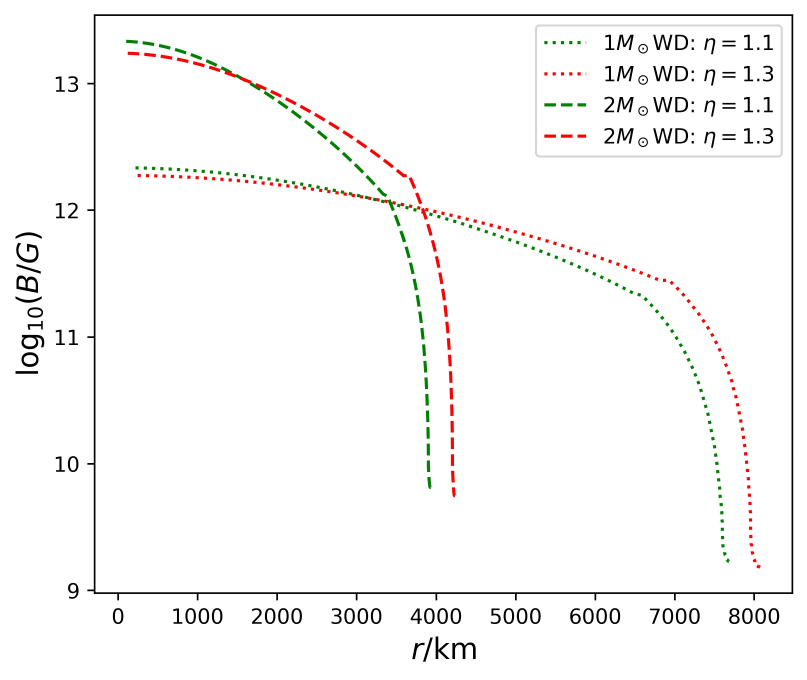}
	\caption{{Variation of magnetic field ($B$) as a function of radius within WD ($r$) for $M =1M_\odot$ and $M =2M_\odot$. In each case, two different $B$ profiles are considered, i.e. $\eta = 1.1$ and $\eta = 1.3$, with $\gamma = 0.8$ throughout. $B_s$ depends on the conserved surface magnetic flux, taken to be $B_sR^2 = 10^5 \ GR_\odot^2$ in all four cases.}}
    \label{Bprofs}
\end{figure}
    
    \item {\textit{Neglecting the effect of rotation and $\alpha$ on stellar structure}}: In the present context, we are concerned with the population of SRPs/LPTs. By definition, these are \textit{slow} rotators. The rotational periods of these sources are much higher than the Keplerian periods of WDs ($\simeq 1-10$ s) and, hence, their effect on the stellar structure is minimal.
    
    
    We can estimate the evolution of the angular velocity $\Omega$ due to the accretion onto the WD, i.e., the spin-up using angular momentum conservation, i.e., $I\Omega = $ constant. Here, $I$ is the moment of inertia of the star. Considering the period of the MS to be same as the period of the Sun, i.e., $P = 609.12$ hours, its evolution as a function of $M$ during accretion is shown in Fig.~\ref{PM}. 

    \begin{figure}[!h]
	\centering
    \includegraphics[scale=0.6]{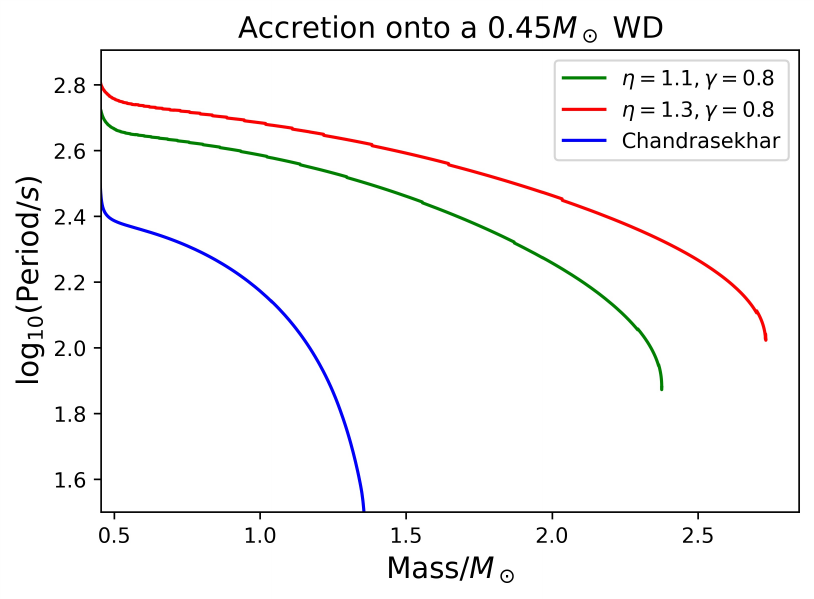}
	\caption{{Variation of $P$ as a function of $M$, showing the increase of the WD's spin period as a result of accretion, for non-magnetic (Chandrasekhar) case as well as different magnetic field profiles. The angular momumentum is conserved during accretion, assuming it is the same as the Sun.}}
    \label{PM}
\end{figure}
    
    As the figure shows, the periods of the magnetized cases remain above the Keplerian frequency throughout. In the case of $\eta = 1.1, \gamma = 0.8$, the saturation $P \simeq$ 76 s, while it is 107 s for $\eta = 1.3, \gamma = 0.8$. The former is close to the shortest rotation period we consider, i.e., for J0901-4046. In summary, for the magnetized cases we consider, rotation can be neglected. Note that depending on the spin period of the initial MS, the exact saturation $P$ may differ, however, Fig. \ref{PM} firmly establishes that (a) the rotation periods in Table~\ref{wdp_table} are achievable by typical accreting magnetized WDs, and (b) the periods remain above the Keplerian limit throughout, i.e., they do not affect stellar structure.
    
    The effect of $\alpha$, i.e. the misalignment betweeen the rotation and magnetic axes on the stellar structure is less  understood, and requires explicit 3D non-axisymmetric codes to be studied. This is beyond the scope of the current work. We expect, again, that the effect of $\alpha$ would be subdominant in its contribution to overall stellar structure.
\end{enumerate}

Further caveats regarding accretion (e.g. compositional effects) have been discussed in detail elsewhere~\citep{ZuraiqApJL}. 

\subsection{Pulsar emission}
The rate of electromagnetic energy release, from a pulsar of angular velocity $\Omega$, follows \citep{LK2004}
\begin{align}
   \dot{E}_{em}&=(2/3c^3)|\boldsymbol{\rm{m}}|^2\Omega^4\sin^2\alpha,
\label{dipole}
\end{align} 
where $|\boldsymbol{\rm{m}}|$ is the magnetic dipole moment and $c$ is the speed of light.

If we assume that there is $100\%$ of rotational energy converted to magnetic dipole radiation, the release of electromagnetic energy is equated with the spin-down luminosity $|\dot{E}_{rot}| = I\Omega\dot{\Omega}$, where $I$ is the moment of inertia of the star. This results in a pulsar braking index\footnote{The braking index ($n$) is defined using the relation $\dot{\Omega} \propto \Omega^n$.}, $n = 3$. In a realistic situation, there may be other conversions at play, e.g. pulsar wind, gravitational waves, making $n \neq 3$. However, $n \neq 3$ does not affect the current value of $B_s$ required to explain pulsar emission. It only affects its evolution over time. We neglect the effect of $n \neq 3$ in the present work.

Thus, the surface dipole magnetic field estimate from eq. \eqref{dipole} is given by 
\begin{equation}
   B_s = \sqrt{\frac{3c^3}{8\pi^2}\frac{I}{R^6\sin^2\alpha}P\dot{P}}.
\label{dipoleB}
\end{equation}

Under the dipole model of pulsar emission, the characteristic age of the source can be computed as (assuming birth period $P_0 << P$),

\begin{eqnarray}
	T=\frac{P}{2\dot{P}}.	
	\label{charage}
\end{eqnarray}

This is a simple model, where we consider the spin-down energy release to be completely driven by dipole surface fields. In reality, pulsars/WDs may have more complicated field geometries, including multipolar field structures \citep{multipole1,multipole2}. We present a general emission formalism that takes into account non-dipolarity as well as deviations in the braking index $n$ in Paper I, and apply it to J0901-4046.
\section{\texttt{STARS} simulated white dwarf pulsar models}
\label{sec:alpha}

We start with examining what the conventional pulsar emission idea, i.e. the emission from dipolar magnetic fields, predicts for the fields of the currently studied sources, for two extreme obliquity angles, viz., $\alpha = 5\degree$ and $\alpha = 90\degree$. Note, $\alpha$ is largely unconstrained for most pulsar sources from observations. 
\begin{table*}[]
\centering
\begin{tabular}{llll}
\hline
                                                                                     & $\alpha$                                        & Source        & $B_s$/G               \\ \hline
\multirow{6}{*}{\begin{tabular}[c]{@{}l@{}}WD \\ ($1M_\odot$, 8000 km)\end{tabular}} & \multicolumn{1}{c}{\multirow{3}{*}{5$\degree$}} & J0901-4046    & $2.11 \times 10^9$    \\
                                                                                     & \multicolumn{1}{c}{}                            & GLEAM-X J1627 & $4.14 \times 10^{11}$ \\
                                                                                     & \multicolumn{1}{c}{}                            & GPM J1839-10  & $1.11 \times 10^{10}$ \\ \cline{2-4} 
                                                                                     & \multirow{3}{*}{90$\degree$}                    & J0901-4046    & $1.84 \times 10^8$    \\
                                                                                     &                                                 & GLEAM-X J1627 & $3.61 \times 10^{10}$ \\
                                                                                     &                                                 & GPM J1839-10  & $9.71 \times 10^8$    \\ \hline
\multirow{6}{*}{\begin{tabular}[c]{@{}l@{}}NS\\ ($1.4M_\odot$, 10 km)\end{tabular}}  & \multirow{3}{*}{5$\degree$}                     & J0901-4046    & $1.60 \times 10^{15}$ \\
                                                                                     &                                                 & GLEAM-X J1627 & $3.13 \times 10^{17}$ \\
                                                                                     &                                                 & GPM J1839-10  & $8.44 \times 10^{15}$ \\ \cline{2-4} 
                                                                                     & \multirow{3}{*}{90$\degree$}                    & J0901-4046    & $1.39 \times 10^{14}$ \\
                                                                                     &                                                 & GLEAM-X J1627 & $2.79 \times 10^{16}$ \\
                                                                                     &                                                 & GPM J1839-10  & $7.35 \times 10^{14}$ \\ \hline
\end{tabular}
\caption{Dipole magnetic fields for low ($5\degree$) and high ($90\degree$) obliquity angles, considering typical $M$ and $R$ of WD and NS.}
\label{table:dipoleNSandWD}
\end{table*}

The dipole fields required to power emission for typical WD and NS cases are shown in Table~\ref{table:dipoleNSandWD}. For typical WD parameters ($M = 1M_\odot$, $R = 8000 \, \rm{km}$), we find that the predicted dipole fields $\sim 10^8 - 10^{11} \ G$. If we check with typical NS parameters ($M = 1.4 M_\odot$, $R = 10 \, \rm{km}$) we find that the field is $\sim 10^{14} - 10^{17} \ G$. 

Across obliquity angles, the NS fields are always above the Schwinger limit, where pair production and pulsation is believed to be suppressed \citep{BH1998} for all the sources. The most magnetized NSs observed today are believed to be the magnetars, with inferred surface fields $\lesssim 10^{15} \ G$, and which are largely a radio-quiet class of objects. Thus, we find that considering a NS origin for these sources goes against current NS observations as well as the understanding of conventional rotation-powered pulsar emission mechanisms.

In the low obliquity angle ($\alpha = 5\degree$) case, it appears that the observed WDs and NSs are on equal footing in terms of explaining the properties of these sources. The fields required to explain the sources are on the higher end and/or exceed the inferred fields from the observed WD/NS. We obtain the highest fields when we consider GLEAM-X J1627. For this source particularly, we obtain fields that are at least two orders of magnitude higher than the typical inferred fields from observed WDs and NSs. As expected, if we now change $\alpha$ from $5\degree$ to $90\degree$, we obtain roughly an order of magnitude decrease in $B_s$ values. The predicted WD fields now range in $10^8 - 10^{10} \, G$, which is within observed WD magnetic field values \citep{ferrario_wd, ferrario2020}. In the NS scenario, we once again see that all fields are above the Schwinger limit, and GLEAM-X J1627 continues to require fields that are well above the observations of magnetars.

\subsection{Exploration of \texttt{STARS}-simulated white dwarfs}
We now turn our attention to the \texttt{STARS}-simulated WDs. We constrain the exact value of $\alpha$ required to explain the slow-rotating sources, as we observe them today, based on WDs with realistic $M$, $R$, and $B_s$ formed after stellar evolution using \texttt{STARS}. The different $R$, $M$ and $B$ are obtained from different magentized stellar evolution paths, differing based on conserved surface flux, poloidal to toroidal percentage, and magnetic field profile.

In order to combine observations of slow-rotating sources with our simulated WDs, we calculate the dipole magnetic field inferred from each of these sources based on standard pulsar physics. We compare the $B_s$ obtained from \texttt{STARS} simulated WDs with the $B_s$ computed from pulsar physics, using eq. \eqref{dipoleB}. We substitute the corresponding stellar $M$ and $R$ from \texttt{STARS}, and vary $\alpha$ between $1\degree$ and $90\degree$ such that the two independent estimates of $B_s$ are equal within a $5\%$ margin of error. We show the lowest and highest magnetic fields of the simulated WDs that can support pulsar emission, along with their corresponding $M, R$ and $\alpha$ in Table \ref{table:alpha}.

\begin{table}[!htbp]
\centering
\begin{tabular}{lllll}
\hline
Source        & $M/M_\odot$ & $R/\rm{km}$ & $\alpha$    & $B_s$/G            \\ \hline
J0901-4046    & 0.80        & 10660       & $12\degree$ & $4.40 \times 10^8$ \\
J0901-4046    & 1.35       & 7238        & $8\degree$ & $1.90 \times 10^9$ \\
GPM J1839-10 & 0.65        & 13657       & $30\degree$ & $5.36 \times 10^8$ \\ 
GPM J1839-10 & 1.35       & 7238        & $46\degree$ & $1.90 \times 10^9$ \\ 
GPM J1839-10 & 1.39       & 6223        & $8\degree$ & $1.29 \times 10^{10}$ \\
GLEAM-X J1627 & 0.65 & 11686 & 69\degree & $1.46 \times 10^{10}$\\ 
GLEAM-X J1627 & 0.82 & 9381 & $90\degree$ & $2.27 \times 10^{10}$\\\hline
\end{tabular}
\caption{A few of the representative \texttt{STARS} simulated WD parameters where the magnetic field was sufficient to explain the corresponding source $P$ and $\dot{P}$.}
\label{table:alpha}
\end{table}

We find that J0901-4046 can be explained by our WDs with \textit{low obliquity angles}. The corresponding $\alpha$ varies between $6\degree$ to  $16\degree$ in various cases. The low obliquity angle is consistent with previous modelling of this source based on its duty cycle that indicated a similarly low $\alpha = 10\degree$ \citep{J0901_10deg}. As shown in the table, the magnetic fields vary between $4 \times 10^8$ to $2\times 10^9 \, G$, consistent with observed WD surface fields \citep{ferrario2020}. In the case of GPM J1839-10, we find that we require higher obliquity angles, i.e., ranging from $33\degree$ to  $50\degree$, if we consider $B_s \approx 10^8 - 10^9 \ G$. However, if we allow $B_s$ to reach $10^{10} \ G$, we find GPM J1839-10 could also be explained with a low $\alpha$ system.

GLEAM-X J1627, however, remains an exception. Within our dataset, simulated WDs with conserved surface flux values up to $5\times 10^5 \ GR_\odot^2$ do not show any result. This is due to its unusually high $P$ and $\dot{P}$, leading to the requirement of high dipolar fields under the conventional pulsar emission model. 

If we consider even higher magnetized cases, i.e. WDs with $B_s \sim 10^{10} G$, corresponding to evolution that conserves a surface flux of $B_sR^2 = 2 \times 10^6 \ GR_\odot^2$, we find a few solutions for GLEAM-X J1627. Crucially, this is still in the range of fluxes observed from magnetized Ap/Bp stars \citep{W2014}. These are the last two rows of Table \ref{table:alpha}. As seen from the table, the sources are low mass, high radius objects with high $
\alpha$. High radius ($\approx 10000$ km) ensures that the sources are dipolar emitters with fields of $\approx 10^{10} \ G$.

However, WDs of $B_s \sim 10^{10} \ G$ are exceedingly rare. It could be that GLEAM-X J1627's emission is not driven by dipolar fields, and instead involves more complicated emission mechanisms, e.g. by multipolar fields. We have presented a generalization of spin down to non-dipolar emission in Paper I. In such a situation, the higher order fields, e.g. quadrupolar components, may be localized and higher in magnitude than the dipole field, leading to more efficient emission. This is in line with similar proposals made for other sources, e.g. GCRT 1745-3009 ($P = 77.13$ min) where emission was proposed to arise from localized, multipolar field structures on the surface of an isolated WD \citep{GilZhang05}. 

\subsection{Death lines}

In order to support pulsar emission, the source must lie above the \textit{death line} in the $P-\dot{P}$ plane. The death line is not unique and there are multiple models depending on the exact model of pulsar emission, i.e., source of pair production, polar cap acceleration model, magnetic field geometry, etc. In the current work, we use the four death lines, defined as multipolar in previous work \citep{bing}. These span the following cases: pair production by either curvature radiation or resonant ICS; and considering either the vacuum gap model or the space-charge–limited flow model for the inner gap model in each case. 

\begin{figure}[!h]
	\centering
    \includegraphics[scale=0.3]{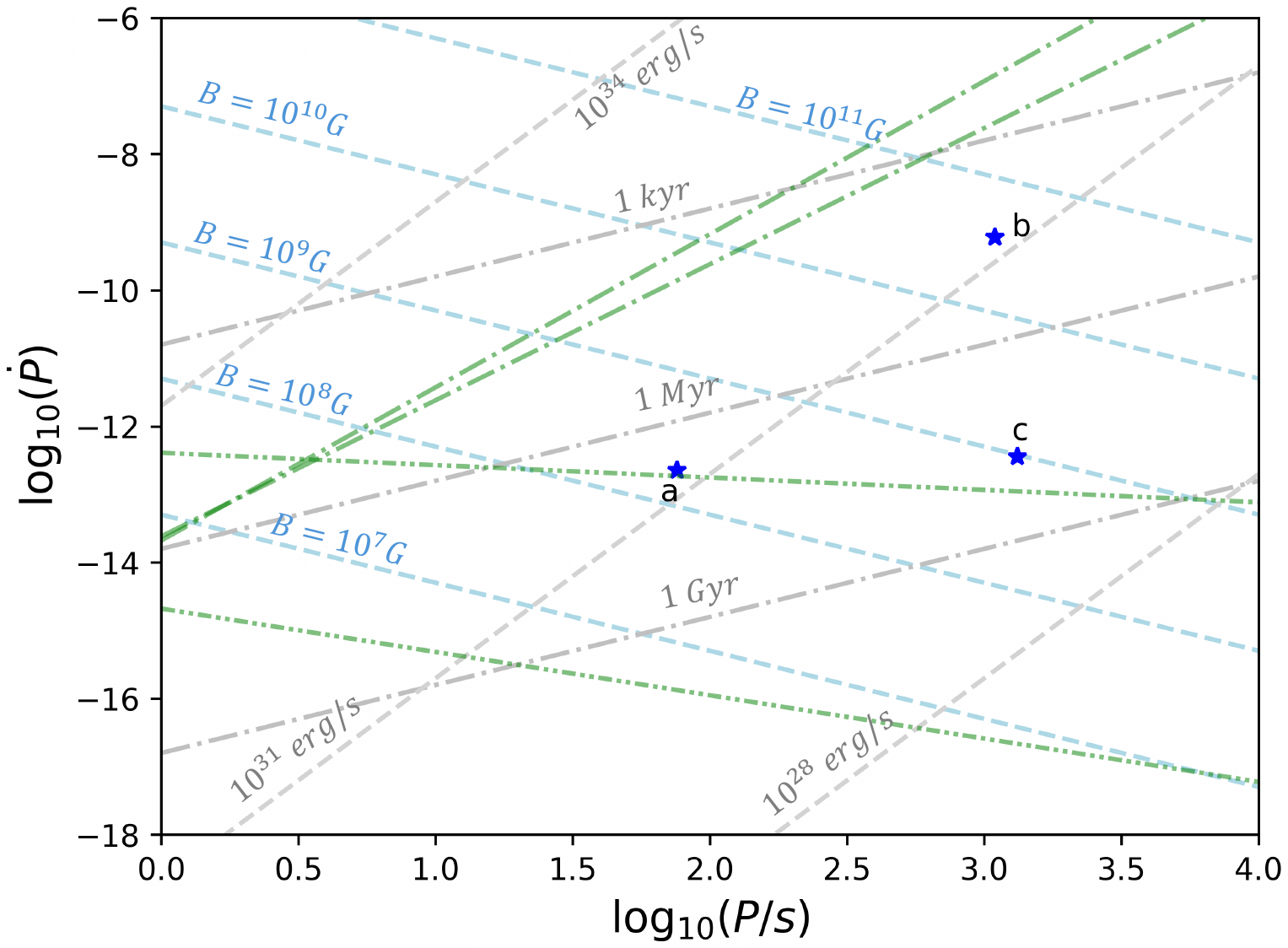}
	\caption{{Constant field lines along with the corresponding  constant age and constant luminosity lines on the $P$-$\dot{P}$ plane. We consider typical WD, $M = M_\odot, R = 8000$ km, $\alpha = 90\degree$. The death lines are also given: the dash-dotted cases show curvature radiation driven pair production, while the dash-double dotted cases show ICS driven pair production. Sources (a) - (c) are labelled as per Table \ref{wdp_table}.}}
    \label{deathlines}
\end{figure}

The positions of our sources on the $P$-$\dot{P}$ plane, along with the death lines, are given in Fig. \ref{deathlines}. For simplicity, we consider the typical WD parameters, i.e., $M = M_\odot$, $R= 8000$ km. We set $\alpha = 90\degree$, i.e., these results show the lower limit of the magnetic field for each source. The corresponding constant age and constant luminosity lines are also shown here.

We see that our pulsar cases do lie above at least some of the death lines we have considered currently, particularly for (resonant) ICS driven pair production. As mentioned above, previous work explored non-resonant ICS as a source of pulsar emission, particularly for WDs \citep{GilZhang05,Zhang1997}, in which case the death lines may be lower. Thus, the currently adopted death lines are a more strict constraint on pulsar properties. From our analysis as well, we see that ICS based death lines (dash-double-dotted in Fig. \ref{deathlines}) are the ones more favored for WD pulsar emission. More discussion of death line constraints and the comparison between death line models for WDs and NSs can be found in Paper I.

These death lines are radius dependent and, hence, are different between WD and NS cases. It is possible that the exact coefficients of the various terms in the death line expressions would be slightly different from those derived in the original work \citep{bing}, owing to differences in WD and NS physics. However, we expect that this change may lie within the range of death lines already considered within the work and, thus, do not expect this to significantly affect the results. 

\section{Super-Chandrasekhar white dwarfs}
\label{sec:SCWD}

An additional intriguing possibility is that of highly magnetized, massive super-Chandraskehar WDs. These are inferred from observations of peculiar, over-luminous type Ia supernovae (SNe Ia). We have found that accretion onto magnetized WDs can lead to strengthening of their central fields, leading to support from the magnetic pressure along with the degeneracy pressure. This leads to the massive WDs that have $M > 1.4M_\odot$, i.e., the super-Chandrasekhar WDs. These sources, being highly magnetized WDs, can serve as a more exotic scenario to explain the slow-rotating sources discussed so far.

In our simulation, we find super-Chandrasekhar WDs formed as a result of accretion on to the dormantly magnetized WDs formed at the end of stellar evolution. This result is also obtained independently in the theoretical model of accretion-based evolution of magnetized, rotating WDs studied in Paper I, as well as previous work \citep{BM2017}. Enhancement of magnetic field in the core of the star due to accretion leads to higher magnetic pressure support and, hence, more massive WDs. Indeed, magnetized WDs as a whole have a higher average mass when compared to their non-magnetized counterparts \citep{ferrario2020}. This is shown in Fig.~\ref{HR}, when we compare dotted lines with the dash-dotted lines. Thus, the presence of magnetic fields in the star can significantly affect stellar evolution and structure, particularly in the later stages of the star's life. See, \cite{ZuraiqApJL} for more details.

The prospect of super-Chandrasekhar WDs is particularly relevant in relation to SRPs/LPTs, since it is expected that such objects would be rare. Peculiar overluminous SNe Ia, thought to arise from this class of objects, are $\lesssim 1\%$ of the total SNe Ia population \citep{sn1a}. Thus, isolated WD pulsars, if they end up being super-Chandrasekhar in mass, could be a much rarer and unusual class of objects when compared to the conventional WD/pulsar populations. Note that even if isolated now, the WD could have been the result of binary evolution \citep{dwd_gleamx, BM2017}.

One of the common arguments against J0901-4046 being a WD pulsar is that it has no detected companion \citep{j0901-long}, compared to AR Scorpii and other pulsars/LPTs which have pulsation arising from binary interactions of a WD with a companion, typically a M-dwarf star.
However, the long-term evolution of J0901-4046 \citep{j0901-long} shows differences in properties from typical magnetar models, further complicating things. 

J0901-4046 being an isolated super-Chandrasekhar WD is then still a viable possibility. It could explain why this source particularly pulses as an isolated WD, when other sub-Chandrasekhar WDs do not - it is simply due to the unusual/rare nature of super-Chandrasekhar and highly magnetized WDs. As shown from our stellar evolution simulations and theoretical calculations, mass transfer from binary partners can lead to WDs growing to super-Chandrasekhar masses. This companion could be too small to be detected at present \citep{parsons26,belloni23}. The enhanced mass of super-Chandrasekhar WD could also be as a result of mass transfer in a system of merging WDs (double degenerate scenario).  
In such cases, we do not expect a companion to be left behind. We have investigated the simulated mass transfer and links to observed binary/accreting WD systems elsewhere~\citep{ZuraiqApJL}. 
On the whole, the absence of a binary partner/companion, particularly when taking the example of J0901-4046, does not rule out a WD origin 
when we take into account the rare/unusual possibility of super-Chandrasekhar WDs. Note that the detection of binary companions has not been confirmed for the other two sources as well. 

Keeping the above-mentioned arguments in mind, we investigate the possibility for super-Chandrasekhar WDs as the possible candidates for our list of sources. The matching was done with the same procedure as in the previous section. The results are given in Table~\ref{table:SCWD}. 

\begin{table}[!h]
\centering
\begin{tabular}{lllll}
\hline
Source & $M/M_\odot$ & $R/\rm{km}$ & $\alpha$    & $B_s$/G            \\ \hline
J0901-4046 & 1.62        & 7068        & $17\degree$ & $1.02 \times 10^9$ \\
J0901-4046 & 2.04       & 4355        & $10\degree$ & $5.27\times 10^9$ \\ 
J0901-4046 & 2.05       & 4317       & $2\degree$ & $2.68 \times 10^{10}$ \\ 
GPM J1839-10 & 1.80      & 6363       & $56\degree$ & $2.47 \times 10^{9}$ \\ 
GPM J1839-10 & 2.04       & 4355        & $62\degree$ & $5.27\times 10^9$ \\ 
GPM J1839-10 & 2.15     & 4000     & $11\degree$ & $3.13\times 10^{10}$ \\ 
GLEAM-X & -    & -    & - & - \\ \hline
\end{tabular}
\label{table:SCWD}
\caption{Same as Table \ref{table:alpha} but for super-Chandrasekhar WDs.}
\end{table}

If we consider super-Chandrasekhar WDs to have surface fields that are of the order of currently observed WDs, i.e., $10^8 - 10^9 \ G$, the results largely follow their sub-Chandrasekhar counterparts from Table \ref{table:alpha}. J0901-4046 requires low obliquity angles , i.e., $\alpha < 20\degree$, while GPM J1839-10 requires a higher $\alpha$. On the other hand, if we allow $B_s \sim 10^{10} \ G$, both J0901-4046 and GPM J1839-10 require a much smaller $\alpha$ to explain emission. This is consistent with the higher field. In general, due to their smaller size, super-Chandrasekhar WDs are expected to have higher surface fields when compared to their sub-Chandrasekhar counterparts. Thus, $10^{10} \ G$ and the corresponding small $\alpha$ may be realistic in this scenario. Surface field $\sim 10^{10}\ G$ is also consistent with our exploration in Paper I, where we have considered accretion-based evolution for J0901-4046, conserving magnetic flux and angular momentum with various obliquity angles. Comparing the various stellar evolution pathways of the current paper and Paper I confirms that the interplay between the field geometry and the obliquity angle is what ultimately determines $B_s$.

However, we do not find any solutions for GLEAM-X J1627. As mentioned previously, the unusually high $\dot{P}$ along with its long period requires high dipole fields to explain emission. In order to have fields not exceeding $10^{10}\ G$, the star must be sufficiently large in radius. Thus, it is the lower mass, higher radii cases which are favoured for a dipole emission model of GLEAM J1627, as shown in Table \ref{table:alpha}. At super-Chandrasekhar WD radii, the field exceeds $10^{11} \ G$. Thus from the analyses of the last two sections, it appears that GLEAM-X J1627 necessarily requires high field and high radii systems to explain its emission based on dipolar spin down.

\section{Conclusions}
\label{sec:conclusions}

Our analysis of the three SRP/LPT sources given in Table~\ref{wdp_table}, with inputs from \texttt{STARS} simulations, leads to the following conclusions:

\begin{itemize}
    \item In the conventional model of dipolar emission from NSs, we find the required fields are as high, or even much higher than their currently observed maximum surface fields. Further, $B_s > B_c$ across all the currently studied sources. This goes against our understanding of pulsar emission~\citep{BH1998}. In the WD case, this is only true for GLEAM-X J1627. Nevertheless, in all WD cases, $B_s <B_c$. 

    \item J0901-4046 has largely been interpreted as a NS in the literature \citep{caleb}. However, as we mention throughout the current work, it cannot completely follow conventional dipolar emission as its predicted $B_s > B_c$, even for $\alpha =90\degree$. When we combine this with pulse width studies that indicate small $\alpha$ \citep{J0901_10deg}, it appears that alternate models should still be pursued. Further, it  does not have a confirmed binary companion yet, and timing studies seem to rule out that scenario \citep{j0901-long}. J0901-4046, thus, could be a good candidate for an isolated WD pulsar. The rarity of such sources also prompts us to consider exotic scenarios for this source, including the possibility of isolated super-Chandrasekhar WDs. Across all the WD cases considered here, i.e. both with sub- and super-Chandrasekhar masses, we consistently require small $\alpha$ to explain the observed emission. 

    \item A WD origin for GPM J1839-10 is apparently unconstrained by our study, while a NS origin requires $B_s \gtrsim 10^{15} \ G$. 
    In the case of a WD origin, this source seems to fit in multiple scenarios, explained by both sub- and super-Chandrasekhar WDs, with either a combination of high $\alpha$ and lower fields ($B_s \sim 10^8 - 10^9 \ G$) or low $\alpha$ and high field ($B_s \sim 10^{10}\ G$). However, it is worth noting here that alternate scenarios to spin down emission, e.g., a WD-M dwarf system with binary interaction driven emission, have recently been proposed for this source \citep{gpm_nature}. Nevertheless, in the absence of the confirmation of any one mechanism, it is worth exploring constraints from dipolar spin-down as we have done here.

    \item GLEAM-X J1627 is the most interesting source as its possible WD origin is the most strongly constrained by our study. In our sample of simulated WDs, GLEAM-X J1627's emission can only be explained by highly magnetized, large radius WDs with high $\alpha$. We do not find solutions for WDs of fields $B_s \sim 10^8 - 10^9 \ G$. We also do not find solutions when we consider super-Chandrasekhar candidates, which generally have smaller radii. In the absence of an observed companion, GLEAM-X could be a good candidate for an isolated WD pulsar, which has also been proposed in other work \citep{dwd_gleamx,katz_gleamx}. GLEAM-X also shows transient behaviour, with emission detectable over 3 months. This leads to comparison with GCRT J1745-3009 ($P = 77.13~\mathrm{min}$), a transient source modelled by magnetospheric emission from non-dipolar field structures of a WD pulsar \citep{GilZhang05}. It is then possible that GLEAM-X J1627 follows a different emission mechanism entirely, perhaps involving multipolar emission.

\end{itemize}

Super-Chandrasekhar WDs have long been believed to be the progenitors of peculiar, over-luminous SNe Ia \citep{SN2003fg,SN2009dc}. They have not been directly observed, but appear to be a natural consequence of enhanced mass support from magnetic effects on stellar structure \citep{DB2013,deb_wd,zz_stars,ZuraiqApJL}. 
In the current work, we have established the realizability and stability of these objects through full-fledged stellar evolution. Linking the simulated WDs arising from these evolutionary trajectories, both at the sub-Chandrasekhar and super-Chandrasekhar level, to observations of unusual pulsar sources, such as the SRPs/LPTs studied in the present work, allows us an additional way to observe the effects of magnetic fields on WDs, and gives us an additional possible avenue towards the study of super-Chandrasekhar WDs.

The preliminary parametric study we have carried out here thus shows that conventional pulsar emission models may not be the best scenario to explain these sources, particularly GLEAM-X J1627. Recent work \citep{Ferrario2026} has shown the possibility of electron cyclotron maser emission (ECME) due to accretion of interstellar medium plasma onto isolated sources driving pulsar emission, though it comes with its own caveats. ECME has also been explored in other binary LPTs \citep{zhang_ilt}. It is clear that some generalization/alternatives to the conventional pulsar emission model are required as the population of LPTs keeps growing. Generalizing the origin of these sources 
to include WDs as well as NSs, as we have done here, is one such step. Further generalization of emission models must also be explored. Non-dipolar/multipolar emission is one possibility. We have initiated this exploration in Paper I.

In the present work, we have assumed that the enhanced magnetic field of the WDs under consideration is a result of accretion from a companion, after their formation. However, there may be evolutionary routes towards such magnetized WDs, e.g. starting with sufficiently massive and magnetized MSs, other than what is explored here. The stellar evolution method explored presently allows us to work with magnetic WDs which are stable under hydrostatic equilibrium and undergo further realistic cooling with time, but is not a unique path towards these objects. It is unclear whether the detailed evolutionary history plays a significant enough role in the emission properties of the WD as we observe it presently. Nevertheless, in the absence of other realistic, magnetized WD models arising from stellar evolution, these are a good starting point for such a study.

In conclusion, while we may not be able to definitively say any of the currently studied sources are WD pulsars, our parameter survey nevertheless shows us that realistic WDs, arising from stellar evolution simulations, can support pulsar emission and explain current observations. It is crucial to generalize conventional NS dipole-based emission in order to explain these new and unusual observations. Future multi-wavelength observations may be able to break the degeneracy between the different scenarios.

\section*{Acknowledgments}
We thank the referee for useful suggestions that have helped us improve the presentation of the work.
Z.~Z.~is supported by the Prime Minister's Research Fellows (PMRF) scheme, Ref. No. TF/PMRF-22-7307.


\bibliography{refs}{}
\bibliographystyle{aasjournalv7}



\end{document}